\documentclass[sigconf,screen,nonacm]{acmart}

\usepackage{listings}
\usepackage{mathpartir}
\usepackage{caption} 
\usepackage{subcaption}
\usepackage{dsfont}
\usepackage{stmaryrd}
\usepackage{float}
\usepackage{microtype}
\usepackage{balance}

\usepackage{xcolor}
\usepackage{color}
\definecolor{keywordcolor}{rgb}{0.7, 0.1, 0.1}   %
\definecolor{tacticcolor}{rgb}{0.0, 0.1, 0.6}    %
\definecolor{commentcolor}{rgb}{0.4, 0.4, 0.4}   %
\definecolor{symbolcolor}{rgb}{0.0, 0.1, 0.6}    %
\definecolor{sortcolor}{rgb}{0.1, 0.5, 0.1}      %
\definecolor{attributecolor}{rgb}{0.7, 0.1, 0.1} %

\AtBeginDocument{%
  }

\def\lchoice{\mathrel{\mkern8mu  \vcenter{\hbox{$\scriptscriptstyle+$}}%
                    \mkern-17mu{\leftarrow}}}
\newcommand{\dn}[1]{\mathcal{D} \llbracket #1 \rrbracket}

\newcommand{\tld}{{\raise.17ex\hbox{$\scriptstyle\sim$}}}
\newcommand{\li}[1]{\lstinline"#1"}

\setcopyright{acmlicensed}
\acmPrice{15.00}
\acmDOI{10.1145/3605156.3606456}
\acmYear{2023}
\copyrightyear{2023}
\acmSubmissionID{isstaws23ftfjpmain-id3-p}
\acmISBN{979-8-4007-0246-4/23/07}
\acmConference[FTfJP '23]{Proceedings of the 25th ACM International Workshop on Formal Techniques for Java-like Programs}{July 18, 2023}{Seattle, WA, USA}
\acmBooktitle{Proceedings of the 25th ACM International Workshop on Formal Techniques for Java-like Programs (FTfJP '23), July 18, 2023, Seattle, WA, USA}
\received{2023-05-26}
\received[accepted]{2023-06-23}

\begin{document}

\title{Using Rewrite Strategies for Efficient Functional Automatic Differentiation}

\author{Timon Böhler}
\email{timon.boehler@stud.tu-darmstadt.de}
\orcid{https://orcid.org/0009-0002-9964-7367}
\affiliation{%
  \institution{Technical University of Darmstadt}
  \city{Darmstadt}
  \country{Germany}
}
\author{David Richter}
\email{david.richter@tu-darmstadt.de}
\orcid{https://orcid.org/0000-0002-8672-0265}
\affiliation{%
  \institution{Technical University of Darmstadt}
  \city{Darmstadt}
  \country{Germany}
}
\author{Mira Mezini}
\email{mezini@informatik.tu-darmstadt.de}
\orcid{https://orcid.org/0000-0001-6563-7537}
\affiliation{%
  \institution{Technical University of Darmstadt}
  \institution{hessian.AI}
  \city{Darmstadt}
  \country{Germany}
}
\begin{abstract}
Automatic Differentiation (AD) has become a dominant technique in ML. AD frameworks have first been implemented for imperative languages using tapes.
Meanwhile, functional implementations of AD have been developed, often based on dual numbers,
which are close to the formal specification of differentiation and hence easier to prove correct.
But these papers have focussed on correctness not efficiency.
Recently, it was shown how an approach using dual numbers could be made efficient through the right optimizations.
Optimizations are highly dependent on order, as one optimization can enable another.
It can therefore be useful to have fine-grained control over the scheduling of optimizations.
One method expresses compiler optimizations as rewrite rules, whose application can be combined and controlled using strategy languages.
Previous work describes the use of term rewriting and strategies to generate high-performance code in a compiler for a functional language.
In this work, we implement dual numbers AD in a functional array programming language using rewrite rules and strategy combinators for optimization.
We aim to combine the elegance of differentiation using dual numbers with a succinct expression of the optimization schedule using a strategy language.
We give preliminary evidence suggesting the viability of the approach on a micro-benchmark.
\end{abstract}



\maketitle

\section{Introduction}

Training a neural network means optimizing the parameters which control its behavior, with respect to a
loss function. The usually employed optimization algorithms, which are based on gradient descent, require
computing the loss function’s gradient~\cite{Wang20}. This means that we need to differentiate the neural network.
While this could in principle be done by hand, automatic differentiation (AD) allows computing the derivative of
a given program without additional programming effort.
As AD is not restricted to the specific operations used
by typical neural networks, more complex constructs, for example involving control flow, can be employed in
machine learning, as long as the program remains differentiable and has trainable parameters. This approach,
which generalizes from deep neural networks to a broader class of programs, has been called differentiable
programming~\cite{LeCun18}.

Models in differentiable programming operate over nested arrays, or tensors. Hence, commonly
used deep learning frameworks like PyTorch~\cite{Paszke19} or JAX~\cite{Jax18} feature a large number of built-in operations on
tensors. In this work, we are instead interested in array languages which have only few built-in
constructs upon which richer APIs can be constructed, like \~{F}~\cite{Shaikhha17, Shaikhha19} or Dex~\cite{Paszke21}. Our implementation of an
array programming language as a domain specific language (DSL) closely follows the former.

Functional approaches to AD based on dual numbers
can be conceptually simple and have been the basis of correctness proofs~\cite{Mazza21}.
There is a challenge when using this approach for ML: gradient descent needs the full gradient
of the loss function. But dual numbers can only compute
the gradient one entry at a time. The size of the gradient is equal to the number of parameters the model has, so
differentiating over a model with $n$ parameters requires $n$ executions of the differentiated function.

To improve the performance of their dual numbers AD algorithm,
\citet{Shaikhha19} present a set of optimization rules.
Optimizations are highly dependent on order, as one optimization can enable another.
It can therefore be useful to have fine-grained control and explore different schedules for the optimization.
One method expressing compiler optimizations as rewrite rules, whose application can be combined and controlled using strategy languages~\cite{Visser98}.
\citet{Hagedorn20} describes the use of term rewriting and strategies to generate high-performance code in a compiler for a functional language, but they did not model differentation.

\paragraph{Contributions}
In this work, we implement dual numbers AD in a functional array programming language using rewrite rules and strategy combinators for optimization.
We aim to \emph{combine the elegance of differentiation using dual numbers with a succinct expression of the optimization schedule using a strategy language}.
We give preliminary evidence suggesting the viability of the approach on a micro-benchmark.
Our array language is implemented in the Lean programming language and theorem prover~\cite{Moura21} as an
embedded DSL. We use Lean’s dependent types to track the sizes of arrays and indices in the type system,
aiming to prevent out-of-bounds errors.

\section{Language} \label{section:typedEmbedded}

\catcode`\<=\active \def<{
\fontencoding{T1}\selectfont\symbol{60}\fontencoding{\encodingdefault}}
\newcommand{\assign}{:=}
\newcommand{\tmop}[1]{\ensuremath{\operatorname{#1}}}
\newcommand{\nonconverted}[1]{\mbox{}}

\begin{figure}[h]
    \centering
  \begin{subfigure}{\linewidth}
    \[ n \in \mathds{Z} \]
    \begin{align*}
      \alpha, \beta &::= \alpha \to \beta~|~\alpha \times \beta~|~\mathtt{array}_n~\alpha~|~\mathtt{int}~|~\mathtt{fin}_n~|~\mathtt{real}
    \end{align*}
      \caption{Types.}
          \label{fig:typesFODSL}
  \end{subfigure}

  \begin{subfigure}{\linewidth}
\begin{align*}
  \dn{\alpha \to \beta} ~\assign~ & \dn{\alpha} \to \dn{\beta} \\
  \dn{\alpha \times \beta} ~\assign~ & \dn{\alpha} \times \dn{\beta} \\
  \dn{\mathtt{array}_n~\alpha} ~\assign~ & \mathtt{array}_n~\dn{\alpha} \\
  \dn{\mathtt{int}} ~\assign~ & \mathtt{int} \\
  \dn{\mathtt{fin}_n} ~\assign~ & \mathtt{fin}_n \\
  \textcolor{blue}{\dn{\mathtt{real}} ~\assign~} & \textcolor{blue}{\mathtt{real} \times \mathtt{real}}
\end{align*}
    \caption{Dual numbers transformation on types.}
    \label{fig:dualTransformTypes}
  \end{subfigure}
  \caption{Definition of types and dual numbers transformation.}
\end{figure}  

\begin{figure*}[h]
  \begin{subfigure}{.49\textwidth}
    \centering
    \begin{mathpar}
      \inferrule{ }{x_{\alpha} : \alpha}
     \  \inferrule{f : \alpha \to \beta \quad a : \alpha}{f~a : \beta}
     \  \inferrule{b : \beta}{\lambda x_{\alpha} .~b : \alpha \to \beta}
     \  \inferrule{e_1 : \alpha \quad e_2 : \beta}{\mathtt{let}~x_{\alpha} := e_1;~e_2 : \beta}
     \\ \inferrule{c : \mathtt{int} \quad e_1 : \alpha \quad e_2 : \alpha}{\mathtt{if}~c~\mathtt{then}~e_1~\mathtt{else}~e_2}
     \quad \inferrule{f : \alpha \to \mathtt{fin}_n \to \alpha \quad x : \alpha}{\mathtt{ifold}_n~f~x : \alpha}
     \\ \inferrule{n \in \mathds{R}}{n : \mathtt{real}}
     \quad \inferrule{i \in \mathds{Z}}{i : \mathtt{int}}
     \quad \inferrule{i \in \{0,...,n-1\}}{i : \mathtt{fin}_n}
    \end{mathpar}
    \begin{align*}
      \mathtt{mkpair} &: \alpha \to \beta \to \alpha \times \beta \\
      \mathtt{fst} &: \alpha \times \beta \to \alpha \\
      \mathtt{snd} &: \alpha \times \beta \to \beta \\
      \mathtt{geti_n} &: \mathtt{array}_n~\alpha \to \mathtt{fin}_n \to \alpha \\
      \mathtt{build_n} &: (\mathtt{fin}_n \to \alpha) \to \mathtt{array}_n~\alpha \\
      + &: \mathtt{real} \to \mathtt{real} \to \mathtt{real} \\
      * &: \mathtt{real} \to \mathtt{real} \to \mathtt{real} \\
      < &: \mathtt{real} \to \mathtt{real} \to \mathtt{int} \\
      = &: \mathtt{int} \to \mathtt{int} \to \mathtt{int} %
    \end{align*}
    \caption{Terms.}
    \label{fig:syntaxFODSL}
  \end{subfigure}%
\begin{subfigure}{.49\textwidth}
  \centering
\begin{align*}
  \dn{x_{\alpha}} ~\assign~ & x_{\dn{\alpha}}\\
  \dn{e_1 e_2} ~\assign~ & \dn{e_1} \dn{e_2}\\
  \dn{ \lambda x_{\alpha}.~e } ~\assign~ & \lambda x_{\dn{\alpha}}.~\dn{e}\\
  \dn{\mathtt{let}~x_{\alpha} := e_1;~e_2} ~\assign~ & \mathtt{let}~x_{\dn{\alpha}} := \dn{e_1};~\dn{e_2} \\
  \dn{\mathtt{if}~c~\mathtt{then}~e_1~\mathtt{else}~e_2}~\assign~& \mathtt{if}~\dn{c}~\mathtt{then}~\dn{e_1}~\mathtt{else}~\dn{e_2} \\
  \dn{\mathtt{ifold}_n~f~x}~\assign~& \mathtt{ifold}_n~\dn{f}~\dn{x} \\
  \textcolor{blue}{\dn{n : \mathtt{real}} ~\assign~} & \textcolor{blue}{(n, 0)}\\
  \textcolor{blue}{\dn{+} ~\assign~} & \textcolor{blue}{\lambda x~y .~(x_1 + y_1, x_2 + y_2)}\\
  \textcolor{blue}{\dn{*} ~\assign~} & \textcolor{blue}{\lambda x~y .~(x_1 + y_1, x_1 * y_2 + x_2 * y_1)}\\
  \textcolor{blue}{\dn{<} ~\assign~} & \textcolor{blue}{\lambda x~y.~x_1 < y_1}\\
  \dn{i : \mathtt{int}} ~\assign~ & i \\
  \dn{i : \mathtt{fin}_n} ~\assign~ &  i \\
  \dn{\mathtt{mkpair}} ~\assign~ & \mathtt{mkpair} \\
  \dn{\mathtt{fst}} ~\assign~ & \mathtt{fst} \\
  \dn{\mathtt{snd}} ~\assign~ & \mathtt{snd} \\
  \dn{\mathtt{geti}_n} ~\assign~ & \mathtt{geti}_n \\
  \dn{\mathtt{build}_n} ~\assign~ & \mathtt{build}_n \\
  \dn{=} ~\assign~ & = %
\end{align*}

    \caption{Dual numbers transformation on terms.}
    \label{fig:dualTransformTerms}
  \end{subfigure}
  \caption{Definition of terms and dual numbers transformation.}
\end{figure*}

For writing models as well as 
optimization and code generation, we use an intrinsically typed deep embedding.
The definition of the types is found in Figure~\ref{fig:typesFODSL}.
Our language has pairs, functions and length-indexed arrays.
The base types are integers, size-bounded natural numbers and real numbers.

We use a limited form of dependent types: array types carry not only
the type of the array's elements, but also the array's length. This means that, for example,
the type $\mathtt{array}_5 \mathtt{int}$ represents arrays of five integers. Additionally,
we use the type $\mathtt{fin}_n$ for indices which represents integers in the range
\texttt{0..n-1}. To retrieve an element from an array of type $\mathtt{array}_n \mathtt{int}$,
we require the index to be of type $\mathtt{fin}_n$. This is intended to prevent
out-of-bounds array accesses. This is similar to the approach used in the Dex array language~\cite{Paszke21}.

Figure~\ref{fig:syntaxFODSL} shows the terms of the language. The language is intrinsically
typed --- \texttt{Term} carries a parameter representing the type
of the expression. Instead of defining the syntax and the type system separately,
the language's constructs are always typed and creating an ill-typed expression is
impossible.
The typing rules do not use
contexts; instead, each variable is labeled by its type~\cite{Church40}.

The terms are variables, function application, let-bindings,
pair construction and projection, if-then-else, iteration, constants for real numbers, integers and
indices as well as pre-defined operations for array construction and indexing, arithmetic, equality checking
and conversion.
Every variable consists of its name and its type.
The typing rule for function application
expects a function of type
$\alpha \to \beta$ and an argument of type $\alpha$ and yields a term of type $\beta$.
The abstraction case is based on the typing rule for lambda abstractions.
Note that we again have to give both a name and a type for each variable.
\texttt{letin} is similar to \texttt{lam} in that it also binds a variable, except we also
already give the value that the variable should be bound to.
\texttt{mkpair} is used to build pairs
and \texttt{fst} and \texttt{snd} are used to take them apart.
We can perform branching with \texttt{if c then e1 else e2}, evaluating the first branch if the
condition is not zero, and the second otherwise.
\texttt{ifold} allows bounded iteration. The fact that the loop index has
type \texttt{fin n} is important when using it to access an array's elements,
for example when computing the sum of an array.
The array operations \texttt{build}, for constructing arrays, and \texttt{geti},
for accessing elements, are now dependently typed. This means that \texttt{geti}
cannot go out of bounds. The types of the two operations also reveal that they are
essentially conversion functions, where \texttt{build} converts from  
\texttt{fin n \tld > a} to \texttt{array n a} and \texttt{geti} converts back.
There is no \texttt{length} operation; as the type of an array expression
now carries its size, \texttt{length} is superfluous. We also have arithmetic operations.

\section{Automatic Differentiation}

The first step in our implementation of AD is a dual numbers transformation~\cite{Shaikhha19}.
As can be seen in Figure~\ref{fig:dualTransformTypes}, it is structurally recursive and
transforms every occurrence of a real number into a pair of real numbers and leaves other type constructors
unchanged. The idea is that each value is bundled with its derivative with regards to some input,
so that both the normal result of the computation and the derivative are computed at the same time.

The transformation on terms is defined in Figure~\ref{fig:dualTransformTerms}.
Its structure follows the structure of the type transformation in that
most of the cases of the transformation are trivial, except for those referring to real numbers.
Variables have only their type changed.
Similarly, the transformation on function application, $\lambda$-abstraction, if-then-else and iteration 
leaves their structure unchanged and simply recursively applies the transformation on
the subexpressions.
For the cases for addition and multiplication of real numbers, note that
we write $(x, y)$ as shorthand for $\mathtt{mkpair}~x~y$ and $x_1$ and $x_2$ as shorthand for
$\mathtt{fst}~x$ and $\mathtt{snd}~x$.
Other operations are unchanged,
e.g. operations on pairs (like \texttt{fst}), arrays (like \texttt{build}),
or integers (like \texttt{addInt}, \texttt{=} or \texttt{fromInt}).
This corresponds to the intuition that we actually just want
to replace constants and arithmetic operations with their dual numbers
equivalents.

In the case of comparisons like $<$, the transformed version simply retrieves
the primal values from the dual numbers given as input and performs the comparison on those.
As Boolean operators are not differentiable, the perturbations of the input numbers
are simply discarded.

\newcommand{\ra}{\mathtt{Array}_a~\mathtt{real}}
\newcommand{\rb}{\mathtt{Array}_b~\mathtt{real}}
\newcommand{\rn}{\mathtt{Array}_n~\mathtt{real}}
\allowdisplaybreaks
With the dual numbers transformation defined, we now want to use it to compute
the gradient of a given model. The AD operators defined here are inspired
by \citet{Shaikhha19}.
\begin{align*}
  &\mathtt{addZeroes}~(v : \mathtt{Array}_n~\alpha) : \mathtt{Array}_n~(\alpha \times \mathtt{real}) \assign \\
  &\quad \mathtt{build}_{|v|}~(\lambda i.~(v [i], 0))\\
  &\mathtt{zip}~(v_1 : \mathtt{Array}_n~\alpha)~(v_2 : \mathtt{Array}_n~\beta) : \mathtt{Array}_n~(\alpha \times \beta) \assign \\
  &\quad \mathtt{build}_n~(\lambda i.~(v_1 [i], v_2 [i]))
\end{align*}
The helper function \lstinline"addZeroes" transforms an array of real numbers into an array of dual numbers
by using each input number as the primal and setting the perturbation to zero,
while \lstinline"zip" combines two arrays of equal length into one.
\begin{align*}
  &\mathtt{oneHot}_n~i \assign \mathtt{build}_n~(\lambda j.~\mathtt{if}~i = j~\mathtt{then}~1~\mathtt{else}~0)\\
  &\mathtt{lossDiff}~e~x~y~p~\bar{p} : \\
  & \quad (\ra \to \rb \to \rn \to \mathtt{real}) \to \\
  & \quad \ra \to \rb \to \rn \to \\
  & \quad \rn \to \mathtt{real} \assign \\
  &\quad \mathtt{snd}~(\dn{e}~(\mathtt{addZeroes}~x)~(\mathtt{addZeroes}~y)~(\mathtt{zip}~p~\bar{p}))
\end{align*}
\lstinline"lossDiff" computes the directional derivative.
It assumes as its argument a loss function that takes three arguments:
an input from the dataset, the corresponding output and the current parameters of
the model. The loss function then returns a number denoting the difference between
the model's output and the true output.
\begin{align*}
  &\mathtt{oneHot}_n~i \assign \mathtt{build}_n~(\lambda j.~\mathtt{if}~i = j~\mathtt{then}~1~\mathtt{else}~0)\\
  &\mathtt{lossGrad}~e~x~y~p : \\
  & \quad (\ra \to \rb \to \rn \to \mathtt{real}) \to \\
  & \quad \ra \to \rb \to \rn \to \\
  & \quad \rn \assign \\
  & \quad \mathtt{build}_{|p|}~(\lambda i.~\mathtt{lossDiff}~e~x~y~p~(\mathtt{oneHot}_{| p |} i))
\end{align*}
\lstinline"lossGrad" computes the full gradient.
It calls \lstinline"lossDiff" multiple times using a one-hot encoding, given by \lstinline"oneHot".
On each call, one entry of the gradient is computed. This implies a number of
executions equal to the dimension of the parameter vector.

\section{Implementation}

Our implementation is written in Lean, a functional programming language and theorem prover~\cite{Moura21}.
We give a short overview of our implementation.

We define terms as a generalized algebraic data type (\lstinline"inductive").
We will omit most of the cases; they are unsurprising and follow the definition given in Fig. \ref{fig:dualTransformTypes}.

\begin{lstlisting}
inductive Term : Typ → Type
| var (x:String) (a) : Term a
| app : Term (a ~> b) → Term a → Term b
| lam (x:String) (a) {b} : Term b → Term (a ~> b)
\end{lstlisting}

The dual numbers transformation is defined as a recursive function operating on the data type.
\begin{lstlisting}
def diff : Term a → Term a.diff
| var x a => var x a.diff
| app e1 e2 => app e1.diff e2.diff
| lam x _ e => lam x _ e.diff
\end{lstlisting}

Lean also features dependent types, which are types that
depend on values.
One example of a dependent type is \texttt{Fin n}, intuitively the type of numbers
from $0$ to $n-1$. Note that type constructors like \texttt{Fin} are just
functions from types to types, so we can define them the same way we define other functions.
\begin{lstlisting}
def Fin (n : Nat) : Type := {i : Nat // i < n}
\end{lstlisting}
The notation on the right-hand side is akin to set builder notation.
It can be read as referring to the type of
all natural numbers \li{i} such that \li{i < n}. More precisely, it
is the type consisting of pairs where the first element is a number \li{i}
and the second element is a proof of the proposition \li{i < n}.

The fact that Lean is dependently typed allows for a very simple implementation
of size-typed arrays and indices, as can be seen in the definition of (a subset of) the types of our DSL:
\begin{lstlisting}
inductive Typ
| real : Typ
| array : Nat → Typ → Typ
| fin : Nat → Typ
\end{lstlisting}
Both \li{array} and \li{fin} are size-typed by indexing them with a natural number
representing the size.

It may seem that \texttt{Term} is quite limited in its expressivity.
The language is simply typed and size parameters for arrays and indices have to
be constants. We can however represent functions that are polymorphic both over types and
over array sizes by quantifying on the level of the metalanguage, Lean.
Consider, for example, the function \texttt{vectorMap}, which applies a function to every
element of an array.
\begin{lstlisting}
def vectorMap {n : Nat} {a b : Typ} :
  Term (array n a ~> (a ~> b) ~> array n b) :=
lam "v" _ (lam "f" _
  (build' (lam "i" _ (app "f" (geti' (var "v" (array n a)) "i")))))
\end{lstlisting}
It is polymorphic with regards to the size of the array as well as to
the type of the function and the array's elements.
This is represented by a Lean function that takes a number
(the size parameter) and two values of type \texttt{Typ} (the type parameters) as input
and returns an \texttt{Term} expression. This way, polymorphic functions
can be specialized to concrete, monomorphic ones. The concrete types can often be inferred
from the context by the Lean type checker.

\section{Optimization} \label{section:opt}

We want to implement rewrite rules and the following optimization strategies, based on \citet{Visser98}, in code.

\begin{center}
(identifiers) $x$ \quad\quad\quad (rules) $r$
\begin{align*}
\text{(strategies)}~s ::= x~|~r~|~\mathsf{id}~|~\lightning~|~s_1;\;s_2
   ~|~s_1 \lchoice s_2~|~\mu x.~s~|~\diamond (s)
\end{align*}
\end{center}
Strategies can be seen as procedures that try to transform terms and either
succeed, returning a new term, or fail.
First, every rewrite rule can be seen as a strategy that, given a term $t$, succeeds if the
rule can be applied to $t$ at the root (so there is no nondeterminism here, as we
do not apply rules to subterms).
We also have the identity strategy \textsf{id}, which always succeeds, leaving the term unchanged.
On the other hand, the strategy $\lightning$ always fails.
We write $s_1 ; s_2$ to denote the sequential composition of two strategies $s_1$ and $s_2$.
Strategy $s_1$ is applied first, and if it succeeds, $s_2$ is applied to the result.
If either $s_1$ or $s_2$ fails, $s_1 ; s_2$ fails.
Left choice $s_1 \lchoice s_2$ first attempts to apply $s_1$.
If the strategy $s_1$ succeeds, its output is returned is the output of $s_1 \lchoice s_2$. If it fails,
$s_2$ is applied on the term.
We also have the fixed point operator $\mu$, which allows us to define recursive strategies.

We also make use of the following, derived, operation: 
\[\text{repeat}(s) := \mu x.~((s ; x) \lchoice \mathsf{id})\]
$\text{repeat}(s)$ iteratively applies a strategy $s$ as often as possible and stops
once $s$ fails. Note that $\text{repeat}(s)$ can never fail; however it may loop indefinitely.
For example, because id never fails, $\text{repeat}(\mathsf{id})$ does not
terminate on any input term.
As rules are only applied to the root, we need a way to rewrite the subterms
of a given term. This is addressed by the $\diamond$ operator.
A strategy $\diamond (s)$ tries to apply $s$ to exactly one subterm of the given
term and fails if there is no subterm to which $s$ can be applied successfully.

In a functional programming language,
this can be done by representing strategies as functions and combinators as
higher-order functions which take and return strategies. Most of the combinators
defined in this section are based on those of the ELEVATE strategy language~\cite{Hagedorn20},
which is itself inspired by Stratego~\cite{Visser05, Visser98}.

The type of an expressions is \li{Term a} for some \li{a}. What should the type of a strategy look
like? The type \li{Term a → Term a} seems sensible, but it assumes that strategies
always produce an expression. Strategies may, however,  fail.
So an improved type would be \li{Term a → Option (Term a)}, where a value of type
\li{Option (Term a)} can either be \li{none}, representing failure, or
\li{some x} (where \li{x} is of type \li{Term a}), representing success.
In our case, this leaves one issue open. We need to be able to generate fresh variable names
(as part of capture-avoiding substitution, for example). How can we do this in
a purely functional language? The answer is to combine \li{Option} with the
state monad, which allows us to thread a state through our computation.
In this case, the state is a natural number which serves as a counter that
is incremented whenever a new variable is produced. The counter is then used as part of the returned
variable name.

This leads us to the following definitions:
\begin{lstlisting}
def RewriteResult a : Type := Nat → Option (a × Nat)
\end{lstlisting}
The meaning of \li{RewriteResult} is that it represents a computation
which takes a counter value and then either fails or returns an output, together
with a new, possibly increased counter value.
It is a monad, allowing the use of do-notation. \\
A strategy is then a function taking an expression and returning a
\li{RewriteResult}, while preserving the type of the expression.
\begin{lstlisting}
def Strategy : Type :=
  {a:Typ} → Term a → RewriteResult (Term a)
\end{lstlisting}

We can now define a function \li{freshM}, which returns a variable name
based on the current counter and increments said counter:
\begin{lstlisting}
def freshM : RewriteResult String
| i => some ("x" ++ toString i, i+1)
\end{lstlisting}

We now implement the strategy combinators.
First we have \texttt{id}, which takes a term and a
counter and returns both unchanged.
\begin{lstlisting}
def id : Strategy := fun p i => some (p, i)
\end{lstlisting}

Failure $\lightning$ is implemented as a function \li{fail}, which always returns
\li{none}. 
\begin{lstlisting}
def fail : Strategy := fun _ _ => none
\end{lstlisting}

Sequencing $s_1;\;s_2$ is represented as \li{seq s1 s2} (abbreviated
as \li{s1 ;; s2}). The code uses do-notation to first apply strategy
\li{s1} to term \li{p}, and then, on success, \li{s2}.
\begin{lstlisting}
def seq (s1 s2 : Strategy) : Strategy :=
  fun p => do s2 (← s1 p)
\end{lstlisting}

We write left choice $s_1 \lchoice s_2$ as \li{lchoice s1 s2} (abbreviated
as \lstinline"s1 <+ s2"). The implementation uses the \lstinline"<|>" operator, which
takes two computations and returns the result of the left one if it succeeds, 
and that of the right one otherwise.
\begin{lstlisting}
def lchoice (s1 s2 : Strategy) : Strategy :=
  fun p => s1 p <|> s2 p
\end{lstlisting}

We do not introduce a fixed point construct $\mu x.~s$, rather, we define
strategies recursively using Lean's support for recursive definitions.
This can be seen in the definition of $\text{repeat}(s)$.
\begin{lstlisting}
partial def repeat (s : Strategy) : Strategy
| _, p => ((s ;; repeat s) <+ id) p
\end{lstlisting}
Lean requires us to add the \li{partial} keyword before \li{def},
indicating that we cannot guarantee termination.

We now consider traversals, which are functions
that transform strategies to allow us to rewrite subexpressions of the current
expression.
\begin{lstlisting}
def Traversal := Strategy → Strategy
\end{lstlisting}

For each constructor in the language, we define traversals for each subexpression
of that constructor. For the function application constructor \li{app}, which contains
two subexpressions (function and argument),
we need two traversals: \li{function s}, which applies \li{s} to the
first subexpression, and \li{argument s}, which applies it to the second.
If \li{function s} or \li{argument s} are applied to anything other than
a function application, they fail.
\begin{lstlisting}
def function : Traversal
| s, _, app f a => do return app (<- s f) a
| _, _, _ => failure

def argument : Traversal
| s, _, app f a => do return app f (<- s a)
| _, _, _ => failure
\end{lstlisting}
The same way, we define traversals for the other constructors, one for each of their respective subexpressions.

We can now implement the combinator \li{one s}~($\diamond s$), which applies \li{s} to one subexpression.
The implementation given here
is deterministic, as it is biased towards the subexpression on the left.
\li{one} works by trying to apply $s$ to every type of subexpression in order.
\begin{lstlisting}
def one (s : Strategy) : Strategy :=
  function s <+ argument s <+ -- other traversals omitted
\end{lstlisting}

\li{one} by itself only allows us do transform expressions that
are direct subexpressions of the root of the abstract syntax tree. To allow transformations
of more deeply nested expressions, we define the recursive \li{topDown} traversal.
\li{topDown s} first tries to apply \li{s} at the root and if that fails,
recurses into the subexpressions until it finds one expression where \li{s} succeeds.
\begin{lstlisting}
partial def topDown : Traversal :=
  fun s => s <+ one (topDown s)
\end{lstlisting}

Combining \li{topDown} with \li{repeat}, we get \li{normalize s},
which repeatedly applies \li{s} until there is no subexpression left to be
transformed.
\begin{lstlisting}
def normalize (s : Strategy) : Strategy :=
  repeat (topDown s)
\end{lstlisting}

We also define \li{run}, which lets us execute a strategy on a term,
by initializing the variable counter to 0, applying the strategy,
and then discarding the new counter at the end.
\begin{lstlisting}
def run : Strategy → Term a → Option (Term a)
| s, p => Prod.fst <$> s p 0
\end{lstlisting}

\subsection{Efficient AD} \label{section:efficientAD}

Deriving the gradient of a function $f$ via 
forward mode AD involves $n$ computations of the function, where $n$ is the size
of $f$'s input vector. This would appear to make forward mode AD unusable for
training large machine learning models.

To address this, \citet{Shaikhha19} present a set of rewrite rules.
Using these to optimize their programs, they are able to achieve performance
on their benchmarks
that is competitive with or superior to frameworks using reverse mode AD.
We can implement these rules as functions of the \li{Strategy} type,
using pattern matching.

As an example consider the following rule, where constructing an array
and immediately retrieving a single element from it is optimized to a simple
function application.
\begin{lstlisting}
def getBuild : Strategy
| _, geti' (build' e1) e2 => return app e1 e2
| _, _ => failure
\end{lstlisting}
The correctness of the rule follows from the following equality that holds in the semantics (not shown) of our 
DSL.
\[\mathtt{geti}~(\mathtt{build}~e_1)~e_2 \quad \equiv \quad e_1~e_2\]
The intuition is that $\mathtt{build}~e_1$ constructs an array that maps an index $i$ to
$e_1~i$ and therefore, indexing this array with $e_2$ gives $e_1~e_2$.

Shaikhha et al. also use a rule for removing \li{let}-bindings by substituting
the bound variable:
\[\mathtt{let}~x_{\alpha} := e_1;~e_2 \quad \equiv \quad e_1[x_a := e_2]\]
In Lean, it looks like this:
\begin{lstlisting}
def letSubst : Strategy
| _, letin x a e1 e2 => subst y a e1 e2
| _, _ => failure
\end{lstlisting}

This may duplicate work if \li{y} is used multiple times in
\li{e1}. So we use the following strategy, where \li{count (freeVars e1) (y, a)}
returns the number of times that the variable occurs free in \li{e1}.
The strategy \li{letSubstN} only substitutes
if the variable does not occur more often than a given threshold.
We use a treshold of 1. This prevents the substitution from duplicating expressions, which could
lead to an exponential slowdown.

\begin{lstlisting}
def letSubstN (n : Nat): Strategy
| _, letin y a e0 e1 => 
  if count (freeVars e1) (y, a) <= n
  then subst y a e0 e1
  else failure
| _, _ => failure
\end{lstlisting}

Substitution in the lambda calculus is subtle; renaming variables may be
necessary to avoid name captures~\cite{Mimram20}.
We address this by using the "sledge-hammer [sic] approach" described by \citet{Jones02}:
before substitution, rename every bound variable in the term
you are substituting for the variable. This is done using the \li{freshTerm}
strategy, which makes use of the \li{RewriteResult} monad to generate fresh variables
through the \li{freshM} function.
The case for \li{lam} contains the expression \li{replaceVar y x a b'},
which evaluates to the result of replacing each occurrence
of \li{var y a} with \li{var x a} in \li{b'}.

\begin{lstlisting}
def freshTerm : Strategy
| _, var y a => return var y a
| _, lam y a b => do
  let b' ← freshTerm b
  let x ← freshM
  return lam x a (replaceVar y x a b')
| _, app f a =>
  do return app (<- freshTerm f) (<- freshTerm a)
-- omitted letin and mkpair for brevity
| _, e => return e
\end{lstlisting}

\section{Evaluation}

\begin{figure}
    \centering
    \begin{subfigure}{.45\linewidth}
      \centering
      \includegraphics[width=\linewidth]{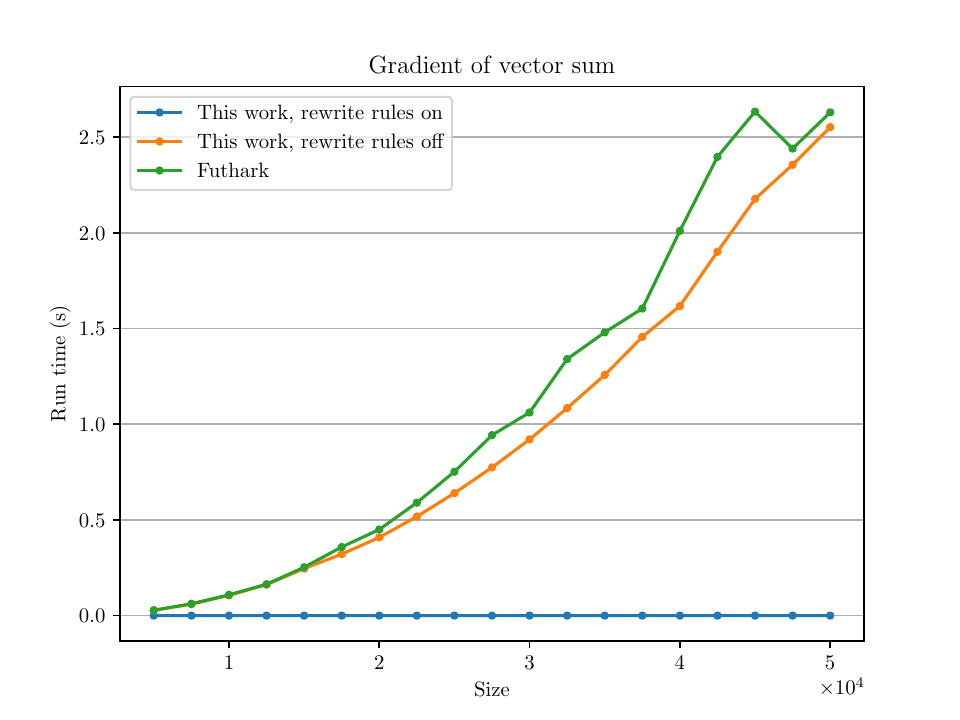}
      \caption{Comparison between the optimized and unoptimized programs.}
      \label{fig:sub1} 
    \end{subfigure} \quad
    \begin{subfigure}{.45\linewidth}
      \centering
      \includegraphics[width=\linewidth]{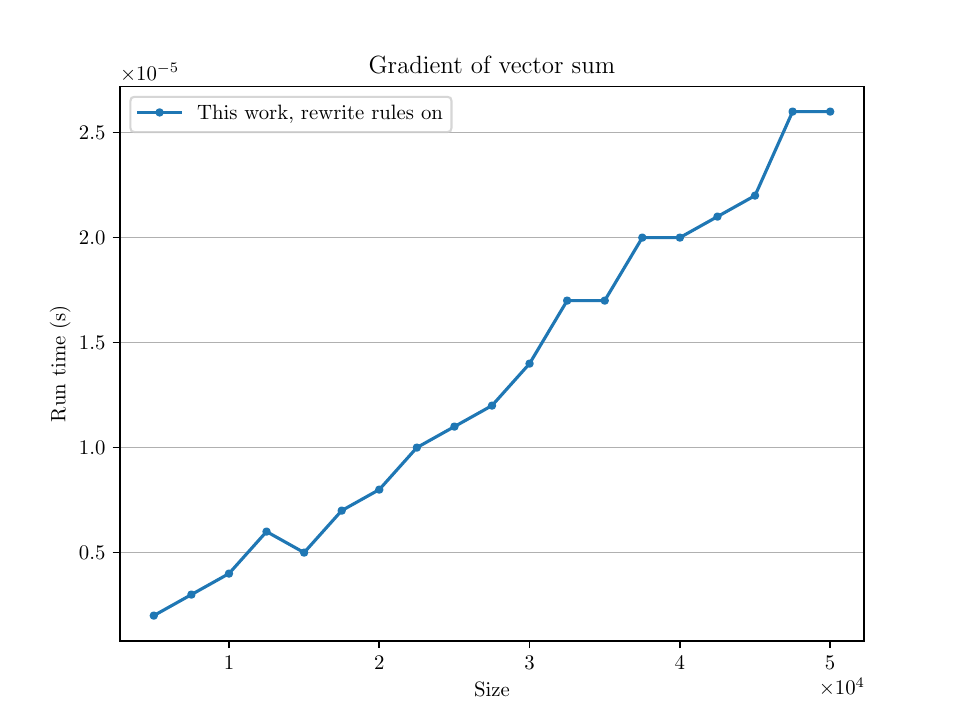}
      \caption{Performance of the optimized program. }
      \label{fig:sub2}
    \end{subfigure}
    \caption{Results of the vector sum benchmark.}
    \label{fig:vecSumBench}
\end{figure}
We conducted a micro-benchmark to test the performance of our implementation
to measure the impact of the optimization rules. 
For the benchmarks we use Python 3.6.9, Futhark 0.22.0, and gcc 7.5.0.
The execution is done on a Intel Pentium G860 (3GHz)
with 4GB of memory.
Our implementation converts the terms to a string representing Futhark \cite{Henriksen17} code, which is
then compiled by the Futhark compiler.
We use the Futhark compiler's C backend.

The micro-benchmark consists of a very simple program which first generates
an array of a given length
where every entry is the same constant value and then computes the 
gradient of \texttt{vectorSum} on that array, where \texttt{vectorSum} is a function
that sums all the entries of an array. This program is somewhat trivial, in that the
\texttt{vectorSum} function is linear and therefore the gradient is always an array
consisting of only 1s. This benchmark should merely demonstrate that the optimizations
from Sec.~\ref{section:efficientAD} can in principle lead to asymptotic speedups.

We tested three different versions. First, a program that is compiled to
Futhark with no optimizations applied before compilation. Second, the same program
with optimizations applied before compilation to Futhark. 
Due to technical issues with compilation, the unoptimized program is generated from the
typed embedding while the optimized one is generated from an untyped one.
This should not affect the qualitative observations we make about the results.
Third, we have a hand-written Futhark program, using a dual numbers library to implement
forward mode AD.

We measured execution time for vector sizes from 2500 to 50000.
The results can be seen in Figure~\ref{fig:vecSumBench}. We give one plot comparing the
three programs and another one focussing only on the runtime of our optimized one.

The left plot shows that the runtime for the unoptimized program in our DSL (orange)
increases faster than linearly. This is expected, as forward mode AD leads to an
overhead proportional to the size of the input vector.
It can be seen however, that the optimized program (blue) is asymptotically faster than the unoptimized one.
The rewrite rules are able to optimize away the nested loops involved in computing the gradient of \texttt{vectorSum}.

Additionally, the hand-written Futhark program (green) is also asymptotically slower than the optimized one.
As all three versions are compiled by the optimizing Futhark compiler, this demonstrates
that we were able to express application-specific optimizations for differentiability in our strategy-based approach which are not included in the
fixed optimization passes of the Futhark compiler.

\section{Related Work}

Forward-mode AD tends to be implemented with dual numbers.
Alternatively, reverse-mode AD allows computing the gradient in one execution, but incurs a complication, as the control flow for computing the
derivative has to be inverted.
Some reverse-mode AD frameworks are implemented as non-compositional transformations,
including Zygote \cite{Innes18b} for Julia, Enzyme \cite{Moses20} for LLVM, and Tapenade \cite{Hascoet13}
for Fortran and C. These lack correctness proofs.
Reverse-mode AD has also been implemented compositionally with continuations \cite{Wang20} or effect handlers \cite{Vilhena21}
as well as mutable state, which may require advanced techniques like separation
logic to verify.
An abstract description of differentiation
is given by categorical models of differentation~\cite{Blute09,Bucciarelli10,Cockett20}.
These do not directly yield an AD algorithm, but can be used to verify one, as has been done by \citet{Cruttwell20}.
Another compositional approach comes from Elliott \cite{Elliott18}, who implements reverse-mode AD
for a first-order language by reifying and transposing the derivative.
\citet{Mazza21} verify the correctness of AD for a Turing-complete higher-order functional language, but use an
inefficient algorithm.
We instead apply a forward-mode transformation and recover efficiency by optimizing the code afterwards,
making use of the flexibility of rewrite strategies.

\section{Conclusion}
We described the implementation of a higher-order functional array language supporting
differentiable programming. Previous work \cite{Shaikhha19},
has not expressed optimizations on differentiated programs using rewrite strategy languages \cite{Visser98, Hagedorn20} and
rewrite strategy languages have not been used for optimizing AD. 
We showed the effect of the optimizations on a micro-benchmark.

\begin{acks}
  This work was funded by
  the German Federal Ministry of Education and Research (BMBF) and the Hessian Ministry of Higher Education, Research, Science and the Arts (HMWK)
  within their joint support of the \textit{National Research Center for Applied Cybersecurity ATHENE}
  and the HMWK via the project 3rd Wave of AI (3AI).
\end{acks}

\balance
\bibliographystyle{ACM-Reference-Format}
\bibliography{base}


\begin{thebibliography}{25}


\ifx \showCODEN    \undefined \def \showCODEN     #1{\unskip}     \fi
\ifx \showDOI      \undefined \def \showDOI       #1{#1}\fi
\ifx \showISBNx    \undefined \def \showISBNx     #1{\unskip}     \fi
\ifx \showISBNxiii \undefined \def \showISBNxiii  #1{\unskip}     \fi
\ifx \showISSN     \undefined \def \showISSN      #1{\unskip}     \fi
\ifx \showLCCN     \undefined \def \showLCCN      #1{\unskip}     \fi
\ifx \shownote     \undefined \def \shownote      #1{#1}          \fi
\ifx \showarticletitle \undefined \def \showarticletitle #1{#1}   \fi
\ifx \showURL      \undefined \def \showURL       {\relax}        \fi
\providecommand\bibfield[2]{#2}
\providecommand\bibinfo[2]{#2}
\providecommand\natexlab[1]{#1}
\providecommand\showeprint[2][]{arXiv:#2}

\bibitem[Blute et~al\mbox{.}(2009)]%
        {Blute09}
\bibfield{author}{\bibinfo{person}{Richard~F Blute}, \bibinfo{person}{J~Robin~B
  Cockett}, {and} \bibinfo{person}{Robert~AG Seely}.}
  \bibinfo{year}{2009}\natexlab{}.
\newblock \showarticletitle{Cartesian differential categories}.
\newblock \bibinfo{journal}{\emph{Theory and Applications of Categories}}
  \bibinfo{volume}{22}, \bibinfo{number}{23} (\bibinfo{year}{2009}),
  \bibinfo{pages}{622--672}.
\newblock


\bibitem[Bradbury et~al\mbox{.}(2018)]%
        {Jax18}
\bibfield{author}{\bibinfo{person}{James Bradbury}, \bibinfo{person}{Roy
  Frostig}, \bibinfo{person}{Peter Hawkins}, \bibinfo{person}{Matthew~James
  Johnson}, \bibinfo{person}{Chris Leary}, \bibinfo{person}{Dougal Maclaurin},
  \bibinfo{person}{George Necula}, \bibinfo{person}{Adam Paszke},
  \bibinfo{person}{Jake Vander{P}las}, \bibinfo{person}{Skye
  Wanderman-{M}ilne}, {and} \bibinfo{person}{Qiao Zhang}.}
  \bibinfo{year}{2018}\natexlab{}.
\newblock \bibinfo{booktitle}{\emph{{JAX}: composable transformations of
  {P}ython+{N}um{P}y programs}}.
\newblock
\urldef\tempurl%
\url{http://github.com/google/jax}
\showURL{%
\tempurl}


\bibitem[Bucciarelli et~al\mbox{.}(2010)]%
        {Bucciarelli10}
\bibfield{author}{\bibinfo{person}{Antonio Bucciarelli},
  \bibinfo{person}{Thomas Ehrhard}, {and} \bibinfo{person}{Giulio Manzonetto}.}
  \bibinfo{year}{2010}\natexlab{}.
\newblock \showarticletitle{Categorical Models for Simply Typed Resource
  Calculi}. In \bibinfo{booktitle}{\emph{Proceedings of the 26th Conference on
  the Mathematical Foundations of Programming Semantics, {MFPS} 2010, Ottawa,
  Ontario, Canada, May 6-10, 2010}} \emph{(\bibinfo{series}{Electronic Notes in
  Theoretical Computer Science}, Vol.~\bibinfo{volume}{265})},
  \bibfield{editor}{\bibinfo{person}{Michael~W. Mislove} {and}
  \bibinfo{person}{Peter Selinger}} (Eds.). \bibinfo{publisher}{Elsevier},
  \bibinfo{pages}{213--230}.
\newblock
\urldef\tempurl%
\url{https://doi.org/10.1016/j.entcs.2010.08.013}
\showDOI{\tempurl}


\bibitem[Church(1940)]%
        {Church40}
\bibfield{author}{\bibinfo{person}{Alonzo Church}.}
  \bibinfo{year}{1940}\natexlab{}.
\newblock \showarticletitle{A Formulation of the Simple Theory of Types}.
\newblock \bibinfo{journal}{\emph{J. Symb. Log.}} \bibinfo{volume}{5},
  \bibinfo{number}{2} (\bibinfo{year}{1940}), \bibinfo{pages}{56--68}.
\newblock
\urldef\tempurl%
\url{https://doi.org/10.2307/2266170}
\showDOI{\tempurl}


\bibitem[Cockett et~al\mbox{.}(2020)]%
        {Cockett20}
\bibfield{author}{\bibinfo{person}{J.~Robin~B. Cockett}, \bibinfo{person}{Geoff
  S.~H. Cruttwell}, \bibinfo{person}{Jonathan Gallagher},
  \bibinfo{person}{Jean{-}Simon~Pacaud Lemay}, \bibinfo{person}{Benjamin
  MacAdam}, \bibinfo{person}{Gordon~D. Plotkin}, {and} \bibinfo{person}{Dorette
  Pronk}.} \bibinfo{year}{2020}\natexlab{}.
\newblock \showarticletitle{Reverse Derivative Categories}. In
  \bibinfo{booktitle}{\emph{28th {EACSL} Annual Conference on Computer Science
  Logic, {CSL} 2020, January 13-16, 2020, Barcelona, Spain}}
  \emph{(\bibinfo{series}{LIPIcs}, Vol.~\bibinfo{volume}{152})},
  \bibfield{editor}{\bibinfo{person}{Maribel Fern{\'{a}}ndez} {and}
  \bibinfo{person}{Anca Muscholl}} (Eds.). \bibinfo{publisher}{Schloss Dagstuhl
  - Leibniz-Zentrum f{\"{u}}r Informatik}, \bibinfo{pages}{18:1--18:16}.
\newblock
\urldef\tempurl%
\url{https://doi.org/10.4230/LIPIcs.CSL.2020.18}
\showDOI{\tempurl}


\bibitem[Cruttwell et~al\mbox{.}(2020)]%
        {Cruttwell20}
\bibfield{author}{\bibinfo{person}{Geoffrey S.~H. Cruttwell},
  \bibinfo{person}{Jonathan Gallagher}, {and} \bibinfo{person}{Dorette Pronk}.}
  \bibinfo{year}{2020}\natexlab{}.
\newblock \showarticletitle{Categorical semantics of a simple differential
  programming language}. In \bibinfo{booktitle}{\emph{Proceedings of the 3rd
  Annual International Applied Category Theory Conference 2020, {ACT} 2020,
  Cambridge, USA, 6-10th July 2020}} \emph{(\bibinfo{series}{{EPTCS}},
  Vol.~\bibinfo{volume}{333})}, \bibfield{editor}{\bibinfo{person}{David~I.
  Spivak} {and} \bibinfo{person}{Jamie Vicary}} (Eds.).
  \bibinfo{pages}{289--310}.
\newblock
\urldef\tempurl%
\url{https://doi.org/10.4204/EPTCS.333.20}
\showDOI{\tempurl}


\bibitem[de~Moura and Ullrich(2021)]%
        {Moura21}
\bibfield{author}{\bibinfo{person}{Leonardo de Moura} {and}
  \bibinfo{person}{Sebastian Ullrich}.} \bibinfo{year}{2021}\natexlab{}.
\newblock \showarticletitle{The Lean 4 Theorem Prover and Programming
  Language}. In \bibinfo{booktitle}{\emph{Automated Deduction - {CADE} 28 -
  28th International Conference on Automated Deduction, Virtual Event, July
  12-15, 2021, Proceedings}} \emph{(\bibinfo{series}{Lecture Notes in Computer
  Science}, Vol.~\bibinfo{volume}{12699})},
  \bibfield{editor}{\bibinfo{person}{Andr{\'{e}} Platzer} {and}
  \bibinfo{person}{Geoff Sutcliffe}} (Eds.). \bibinfo{publisher}{Springer},
  \bibinfo{pages}{625--635}.
\newblock
\urldef\tempurl%
\url{https://doi.org/10.1007/978-3-030-79876-5\_37}
\showDOI{\tempurl}


\bibitem[de~Vilhena and Pottier(2021)]%
        {Vilhena21}
\bibfield{author}{\bibinfo{person}{Paulo~Em{\'\i}lio de Vilhena} {and}
  \bibinfo{person}{Fran{\c{c}}ois Pottier}.} \bibinfo{year}{2021}\natexlab{}.
\newblock \showarticletitle{Verifying an Effect-Handler-Based Define-By-Run
  Reverse-Mode AD Library}.
\newblock \bibinfo{journal}{\emph{arXiv preprint arXiv:2112.07292}}
  (\bibinfo{year}{2021}).
\newblock


\bibitem[Elliott(2018)]%
        {Elliott18}
\bibfield{author}{\bibinfo{person}{Conal Elliott}.}
  \bibinfo{year}{2018}\natexlab{}.
\newblock \showarticletitle{The simple essence of automatic differentiation}.
\newblock \bibinfo{journal}{\emph{Proc. {ACM} Program. Lang.}}
  \bibinfo{volume}{2}, \bibinfo{number}{{ICFP}} (\bibinfo{year}{2018}),
  \bibinfo{pages}{70:1--70:29}.
\newblock
\urldef\tempurl%
\url{https://doi.org/10.1145/3236765}
\showDOI{\tempurl}


\bibitem[Hagedorn et~al\mbox{.}(2020)]%
        {Hagedorn20}
\bibfield{author}{\bibinfo{person}{Bastian Hagedorn}, \bibinfo{person}{Johannes
  Lenfers}, \bibinfo{person}{Thomas Koehler}, \bibinfo{person}{Xueying Qin},
  \bibinfo{person}{Sergei Gorlatch}, {and} \bibinfo{person}{Michel Steuwer}.}
  \bibinfo{year}{2020}\natexlab{}.
\newblock \showarticletitle{Achieving high-performance the functional way: a
  functional pearl on expressing high-performance optimizations as rewrite
  strategies}.
\newblock \bibinfo{journal}{\emph{Proc. {ACM} Program. Lang.}}
  \bibinfo{volume}{4}, \bibinfo{number}{{ICFP}} (\bibinfo{year}{2020}),
  \bibinfo{pages}{92:1--92:29}.
\newblock
\urldef\tempurl%
\url{https://doi.org/10.1145/3408974}
\showDOI{\tempurl}


\bibitem[Hasco{\"{e}}t and Pascual(2013)]%
        {Hascoet13}
\bibfield{author}{\bibinfo{person}{Laurent Hasco{\"{e}}t} {and}
  \bibinfo{person}{Val{\'{e}}rie Pascual}.} \bibinfo{year}{2013}\natexlab{}.
\newblock \showarticletitle{The Tapenade automatic differentiation tool:
  Principles, model, and specification}.
\newblock \bibinfo{journal}{\emph{{ACM} Trans. Math. Softw.}}
  \bibinfo{volume}{39}, \bibinfo{number}{3} (\bibinfo{year}{2013}),
  \bibinfo{pages}{20:1--20:43}.
\newblock
\urldef\tempurl%
\url{https://doi.org/10.1145/2450153.2450158}
\showDOI{\tempurl}


\bibitem[Henriksen et~al\mbox{.}(2017)]%
        {Henriksen17}
\bibfield{author}{\bibinfo{person}{Troels Henriksen}, \bibinfo{person}{Niels
  G.~W. Serup}, \bibinfo{person}{Martin Elsman}, \bibinfo{person}{Fritz
  Henglein}, {and} \bibinfo{person}{Cosmin~E. Oancea}.}
  \bibinfo{year}{2017}\natexlab{}.
\newblock \showarticletitle{Futhark: purely functional GPU-programming with
  nested parallelism and in-place array updates}. In
  \bibinfo{booktitle}{\emph{Proceedings of the 38th {ACM} {SIGPLAN} Conference
  on Programming Language Design and Implementation, {PLDI} 2017, Barcelona,
  Spain, June 18-23, 2017}}, \bibfield{editor}{\bibinfo{person}{Albert Cohen}
  {and} \bibinfo{person}{Martin~T. Vechev}} (Eds.). \bibinfo{publisher}{{ACM}},
  \bibinfo{pages}{556--571}.
\newblock
\urldef\tempurl%
\url{https://doi.org/10.1145/3062341.3062354}
\showDOI{\tempurl}


\bibitem[Innes(2018)]%
        {Innes18b}
\bibfield{author}{\bibinfo{person}{Michael Innes}.}
  \bibinfo{year}{2018}\natexlab{}.
\newblock \showarticletitle{Don't Unroll Adjoint: Differentiating SSA-Form
  Programs}.
\newblock \bibinfo{journal}{\emph{CoRR}}  \bibinfo{volume}{abs/1810.07951}
  (\bibinfo{year}{2018}).
\newblock
\showeprint[arXiv]{1810.07951}
\urldef\tempurl%
\url{http://arxiv.org/abs/1810.07951}
\showURL{%
\tempurl}


\bibitem[Jones and Marlow(2002)]%
        {Jones02}
\bibfield{author}{\bibinfo{person}{Simon L.~Peyton Jones} {and}
  \bibinfo{person}{Simon Marlow}.} \bibinfo{year}{2002}\natexlab{}.
\newblock \showarticletitle{Secrets of the Glasgow Haskell Compiler inliner}.
\newblock \bibinfo{journal}{\emph{J. Funct. Program.}} \bibinfo{volume}{12},
  \bibinfo{number}{4{\&}5} (\bibinfo{year}{2002}), \bibinfo{pages}{393--433}.
\newblock
\urldef\tempurl%
\url{https://doi.org/10.1017/S0956796802004331}
\showDOI{\tempurl}


\bibitem[Mazza and Pagani(2021)]%
        {Mazza21}
\bibfield{author}{\bibinfo{person}{Damiano Mazza} {and}
  \bibinfo{person}{Michele Pagani}.} \bibinfo{year}{2021}\natexlab{}.
\newblock \showarticletitle{Automatic differentiation in {PCF}}.
\newblock \bibinfo{journal}{\emph{Proc. {ACM} Program. Lang.}}
  \bibinfo{volume}{5}, \bibinfo{number}{{POPL}} (\bibinfo{year}{2021}),
  \bibinfo{pages}{1--27}.
\newblock
\urldef\tempurl%
\url{https://doi.org/10.1145/3434309}
\showDOI{\tempurl}


\bibitem[Mimram(2020)]%
        {Mimram20}
\bibfield{author}{\bibinfo{person}{Samuel Mimram}.}
  \bibinfo{year}{2020}\natexlab{}.
\newblock \bibinfo{booktitle}{\emph{PROGRAM = PROOF}}.
\newblock


\bibitem[Moses and Churavy(2020)]%
        {Moses20}
\bibfield{author}{\bibinfo{person}{William~S. Moses} {and}
  \bibinfo{person}{Valentin Churavy}.} \bibinfo{year}{2020}\natexlab{}.
\newblock \showarticletitle{Instead of Rewriting Foreign Code for Machine
  Learning, Automatically Synthesize Fast Gradients}. In
  \bibinfo{booktitle}{\emph{Advances in Neural Information Processing Systems
  33: Annual Conference on Neural Information Processing Systems 2020, NeurIPS
  2020, December 6-12, 2020, virtual}}, \bibfield{editor}{\bibinfo{person}{Hugo
  Larochelle}, \bibinfo{person}{Marc'Aurelio Ranzato}, \bibinfo{person}{Raia
  Hadsell}, \bibinfo{person}{Maria{-}Florina Balcan}, {and}
  \bibinfo{person}{Hsuan{-}Tien Lin}} (Eds.).
\newblock
\urldef\tempurl%
\url{https://proceedings.neurips.cc/paper/2020/hash/9332c513ef44b682e9347822c2e457ac-Abstract.html}
\showURL{%
\tempurl}


\bibitem[Paszke et~al\mbox{.}(2019)]%
        {Paszke19}
\bibfield{author}{\bibinfo{person}{Adam Paszke}, \bibinfo{person}{Sam Gross},
  \bibinfo{person}{Francisco Massa}, \bibinfo{person}{Adam Lerer},
  \bibinfo{person}{James Bradbury}, \bibinfo{person}{Gregory Chanan},
  \bibinfo{person}{Trevor Killeen}, \bibinfo{person}{Zeming Lin},
  \bibinfo{person}{Natalia Gimelshein}, \bibinfo{person}{Luca Antiga},
  \bibinfo{person}{Alban Desmaison}, \bibinfo{person}{Andreas K{\"{o}}pf},
  \bibinfo{person}{Edward~Z. Yang}, \bibinfo{person}{Zachary DeVito},
  \bibinfo{person}{Martin Raison}, \bibinfo{person}{Alykhan Tejani},
  \bibinfo{person}{Sasank Chilamkurthy}, \bibinfo{person}{Benoit Steiner},
  \bibinfo{person}{Lu Fang}, \bibinfo{person}{Junjie Bai}, {and}
  \bibinfo{person}{Soumith Chintala}.} \bibinfo{year}{2019}\natexlab{}.
\newblock \showarticletitle{PyTorch: An Imperative Style, High-Performance Deep
  Learning Library}. In \bibinfo{booktitle}{\emph{Advances in Neural
  Information Processing Systems 32: Annual Conference on Neural Information
  Processing Systems 2019, NeurIPS 2019, December 8-14, 2019, Vancouver, BC,
  Canada}}, \bibfield{editor}{\bibinfo{person}{Hanna~M. Wallach},
  \bibinfo{person}{Hugo Larochelle}, \bibinfo{person}{Alina Beygelzimer},
  \bibinfo{person}{Florence d'Alch{\'{e}}{-}Buc}, \bibinfo{person}{Emily~B.
  Fox}, {and} \bibinfo{person}{Roman Garnett}} (Eds.).
  \bibinfo{pages}{8024--8035}.
\newblock
\urldef\tempurl%
\url{https://proceedings.neurips.cc/paper/2019/hash/bdbca288fee7f92f2bfa9f7012727740-Abstract.html}
\showURL{%
\tempurl}


\bibitem[Paszke et~al\mbox{.}(2021)]%
        {Paszke21}
\bibfield{author}{\bibinfo{person}{Adam Paszke}, \bibinfo{person}{Daniel~D.
  Johnson}, \bibinfo{person}{David Duvenaud}, \bibinfo{person}{Dimitrios
  Vytiniotis}, \bibinfo{person}{Alexey Radul}, \bibinfo{person}{Matthew~J.
  Johnson}, \bibinfo{person}{Jonathan Ragan{-}Kelley}, {and}
  \bibinfo{person}{Dougal Maclaurin}.} \bibinfo{year}{2021}\natexlab{}.
\newblock \showarticletitle{Getting to the point: index sets and
  parallelism-preserving autodiff for pointful array programming}.
\newblock \bibinfo{journal}{\emph{Proc. {ACM} Program. Lang.}}
  \bibinfo{volume}{5}, \bibinfo{number}{{POPL}} (\bibinfo{year}{2021}),
  \bibinfo{pages}{1--29}.
\newblock
\urldef\tempurl%
\url{https://doi.org/10.1145/3473593}
\showDOI{\tempurl}


\bibitem[Shaikhha et~al\mbox{.}(2017)]%
        {Shaikhha17}
\bibfield{author}{\bibinfo{person}{Amir Shaikhha}, \bibinfo{person}{Andrew~W.
  Fitzgibbon}, \bibinfo{person}{Simon~Peyton Jones}, {and}
  \bibinfo{person}{Dimitrios Vytiniotis}.} \bibinfo{year}{2017}\natexlab{}.
\newblock \showarticletitle{Destination-passing style for efficient memory
  management}. In \bibinfo{booktitle}{\emph{Proceedings of the 6th {ACM}
  {SIGPLAN} International Workshop on Functional High-Performance Computing,
  FHPC@ICFP 2017, Oxford, UK, September 7, 2017}},
  \bibfield{editor}{\bibinfo{person}{Phil Trinder} {and}
  \bibinfo{person}{Cosmin~E. Oancea}} (Eds.). \bibinfo{publisher}{{ACM}},
  \bibinfo{pages}{12--23}.
\newblock
\urldef\tempurl%
\url{https://doi.org/10.1145/3122948.3122949}
\showDOI{\tempurl}


\bibitem[Shaikhha et~al\mbox{.}(2019)]%
        {Shaikhha19}
\bibfield{author}{\bibinfo{person}{Amir Shaikhha}, \bibinfo{person}{Andrew~W.
  Fitzgibbon}, \bibinfo{person}{Dimitrios Vytiniotis}, {and}
  \bibinfo{person}{Simon~Peyton Jones}.} \bibinfo{year}{2019}\natexlab{}.
\newblock \showarticletitle{Efficient differentiable programming in a
  functional array-processing language}.
\newblock \bibinfo{journal}{\emph{Proc. {ACM} Program. Lang.}}
  \bibinfo{volume}{3}, \bibinfo{number}{{ICFP}} (\bibinfo{year}{2019}),
  \bibinfo{pages}{97:1--97:30}.
\newblock
\urldef\tempurl%
\url{https://doi.org/10.1145/3341701}
\showDOI{\tempurl}


\bibitem[Visser(2005)]%
        {Visser05}
\bibfield{author}{\bibinfo{person}{Eelco Visser}.}
  \bibinfo{year}{2005}\natexlab{}.
\newblock \showarticletitle{A survey of strategies in rule-based program
  transformation systems}.
\newblock \bibinfo{journal}{\emph{J. Symb. Comput.}} \bibinfo{volume}{40},
  \bibinfo{number}{1} (\bibinfo{year}{2005}), \bibinfo{pages}{831--873}.
\newblock
\urldef\tempurl%
\url{https://doi.org/10.1016/j.jsc.2004.12.011}
\showDOI{\tempurl}


\bibitem[Visser et~al\mbox{.}(1998)]%
        {Visser98}
\bibfield{author}{\bibinfo{person}{Eelco Visser},
  \bibinfo{person}{Zine{-}El{-}Abidine Benaissa}, {and}
  \bibinfo{person}{Andrew~P. Tolmach}.} \bibinfo{year}{1998}\natexlab{}.
\newblock \showarticletitle{Building Program Optimizers with Rewriting
  Strategies}. In \bibinfo{booktitle}{\emph{Proceedings of the third {ACM}
  {SIGPLAN} International Conference on Functional Programming {(ICFP} '98),
  Baltimore, Maryland, USA, September 27-29, 1998}},
  \bibfield{editor}{\bibinfo{person}{Matthias Felleisen}, \bibinfo{person}{Paul
  Hudak}, {and} \bibinfo{person}{Christian Queinnec}} (Eds.).
  \bibinfo{publisher}{{ACM}}, \bibinfo{pages}{13--26}.
\newblock
\urldef\tempurl%
\url{https://doi.org/10.1145/289423.289425}
\showDOI{\tempurl}


\bibitem[Wang et~al\mbox{.}(2018)]%
        {Wang20}
\bibfield{author}{\bibinfo{person}{Fei Wang}, \bibinfo{person}{Xilun Wu},
  \bibinfo{person}{Gr{\'{e}}gory~M. Essertel}, \bibinfo{person}{James~M.
  Decker}, {and} \bibinfo{person}{Tiark Rompf}.}
  \bibinfo{year}{2018}\natexlab{}.
\newblock \showarticletitle{Demystifying Differentiable Programming:
  Shift/Reset the Penultimate Backpropagator}.
\newblock \bibinfo{journal}{\emph{CoRR}}  \bibinfo{volume}{abs/1803.10228}
  (\bibinfo{year}{2018}).
\newblock
\showeprint[arXiv]{1803.10228}
\urldef\tempurl%
\url{http://arxiv.org/abs/1803.10228}
\showURL{%
\tempurl}


\bibitem[{Yann LeCun}(2018)]%
        {LeCun18}
\bibfield{author}{\bibinfo{person}{{Yann LeCun}}.}
  \bibinfo{year}{2018}\natexlab{}.
\newblock \bibinfo{title}{Yann LeCun - OK, Deep Learning has outlived its
  usefulness... | Facebook}.
\newblock
\newblock
\urldef\tempurl%
\url{https://web.archive.org/web/20180106001630/https://www.facebook.com/yann.lecun/posts/10155003011462143}
\showURL{%
\tempurl}
\newblock
\shownote{[Online; accessed 7-April-2022]}.


\end{thebibliography}

\end{document}